\documentclass[aps,preprint,floatfix,nofootinbib,showpacs]{revtex4-1}
\usepackage{graphicx,color,epstopdf,amsmath}
\usepackage{multirow}

\newcommand{\lsim}{\mathrel{\mathop{\kern 0pt \rlap
  {\raise.2ex\hbox{$<$}}}
  \lower.9ex\hbox{\kern-.190em $\sim$}}}
\newcommand{\gsim}{\mathrel{\mathop{\kern 0pt \rlap
  {\raise.2ex\hbox{$>$}}}
  \lower.9ex\hbox{\kern-.190em $\sim$}}}
\newcommand{\met}{ \rlap{\,/}{E}_T}
\newcommand{\gev}{{\;{\rm GeV}}}
\newcommand{\tev}{{\;{\rm TeV}}}

\newcommand{\ttb}{t\bar{t}}
\newcommand{\sg}{{\sigma}}

\newcommand{\kp}{{\kappa}}

\newcommand{\Lm}{{\Lambda}}
\newcommand{\gm}{{\gamma}}

\newcommand{\rr}{{\gamma\gamma}}
\newcommand{\beq}{\begin{equation}}
\newcommand{\eeq}{\end{equation}}
\newcommand{\bea}     {\begin{eqnarray}}
\newcommand{\eea}     {\end{eqnarray}}
\newcommand{\bit}     {\begin{itemize}}
\newcommand{\eit}     {\end{itemize}}

\newcommand{\br}{{\rm BR}}

\newcommand{\pb}{{\,{\rm pb}}}
\newcommand{\fb}{{\,{\rm fb}}}
\newcommand{\ifb}{{\,{\rm fb}^{-1}}}

\newcommand{\bb}{ {b\bar{b}} }

\begin{document}
\title{Using the Higgs boson to probe the littlest Higgs model \\ with $T$-parity 
through $Z_H W_H$ production at the LHC}
\bigskip
\author{Kingman Cheung$^{1,2}$,  Kang Young Lee$^{3}$,  So Young Shim$^{1}$,  
Jeonghyeon Song$^{1}$, and Namseok Yoo$^{1}$}
\affiliation{
$^1$Division of Quantum Phases \& Devices, School of Physics, 
Konkuk University, Seoul 143-701, Korea \\
$^2$Department of Physics, National Tsing Hua University, 
Hsinchu 300, Taiwan \\
$^3$Department of Physics Education, 
Gyeongsang National University,
Jinju 660-701,
Korea
}
\date{\today}

\begin{abstract}
In the littlest Higgs model with $T$-parity,
the production cross section of the $T$-odd heavy gauge boson pair $Z_H W_H$
is sizable at the LHC.
In addition, both the $Z_H$  and $W_H$ bosons have almost exclusively 
one decay channel into $H A_H$ and  $W A_H$  respectively,
where the dark matter candidate $A_H$ yields a large missing transverse energy signal.
Upon the discovery of the Higgs boson at 125 GeV,
we study the discovery sensitivity of the final state 
$ pp \to \ell b\bar{b} +\rlap{\,/}{E}_T$ 
to probe the model at the LHC.
We find that the standard model backgrounds are manageable
by applying suitable kinematic cuts.
The LHC running at $\sqrt{s}=14\tev$ with a $100\ifb$ total luminosity
is sensitive to the model with the signal significance above 5 
if the symmetry breaking scale $f$ is below 
about 850 GeV.
\end{abstract}

\maketitle

\section{Introduction}
\label{sec:introduction}
After a long wait, a particle with close resemblance to the standard model (SM) Higgs boson 
was finally discovered at the Large Hadron Collider (LHC).
Based on 2011 and 2012 data at $\sqrt{s}=7\tev$ and $\sqrt{s}=8\tev$,
the observation of a new boson with mass around 125 GeV 
was declared by the ATLAS and CMS collaborations with a local significance of 
$5.0$ standard deviation~\cite{ATLAS-Higgs,CMS-Higgs,2011:Higgs:ATLAS,2011:Higgs:CMS}.
The CDF and D0 collaborations also supported this discovery through
the excess in the $H \to b\bar{b}$ channel~\cite{Tevatron-Higgs}.
%This new boson 
%shows remarkably similar behaviors to the SM Higgs boson~\cite{HiggsCoupling}.

The excitement at this great discovery is common in the whole particle physics community.
The interpretation of the detailed data splits though.
One approach is to take some deviation of the Higgs data from the SM expectation
as a signal of new physics~\cite{125:NP},
especially the enhanced rate in the $\rr$ channel with about $2\sigma$~\cite{diphoton:excess}.
The other approach is to accept the observed new boson 
as the SM Higgs boson with mass of 125 GeV.
The $2\sigma$ deviation in the $\gm\gm$ mode could fade away with more data.
%Or even the current $H \to \rr$ data can be interpreted as having a mild deviation with $\sim 1\sigma$
%if considering 
%the theoretical uncertainties from scale dependence, 
%from parton distribution functions, and from the use of the effective 
%field theory in calculating the loop contributions
%reduce the deviation in the $\rr$ channel to a mild significance level of 
%~\cite{SM:consistency}.

We take an alternative stance: new physics beyond the SM exists at the TeV scale,
and we probe it by using the SM-like dominant decay of the Higgs boson, 
not by the deviation in the $H\to \gm\gm$ channel.
The hint of new physics is from the observed mass of the Higgs boson,
which demands an answer to the gauge hierarchy problem.
Accepting almost the same properties with the \emph{SM} Higgs boson,
we can use it as a tagging particle for new physics.
This is very useful especially when the Higgs boson is a decay 
product of a heavy new particle.
The final state is a tagged Higgs boson,  \textit{e.g.},
through a $b$ quark pair with the invariant mass near 125 GeV,
accompanied by additional exotic signals such as large missing 
transverse energy $\met$.

In this direction, the littlest Higgs model with $T$-parity invariance,
called the LHT model~\cite{littlest:higgs,Low:JHEP,Cheng:Low},
has drawn a lot of interest as it provides an answer to 
the gauge hierarchy problem and a candidate for the 
cold dark matter.
The model is based on an $SU(5)/SO(5)$ non-linear sigma model,
accommodating the Higgs boson as a pseudo-Nambu-Goldstone boson.
Collective symmetry breaking mechanism prohibits
one-loop quadratic divergence 
in the radiative corrections to the mass of the Higgs boson.
The little hierarchy problem is postponed to much higher energy $O(10)$ TeV.
Phenomenologically, each SM particle has 
a new heavy partner.
In order to weaken the strong constraint from the
electroweak precision data~\cite{EWPD},
$T$-parity was implemented later~\cite{Low:JHEP,Cheng:Low},
under which the new partners of the SM particles have odd parity
while the SM particles have even parity.
The lightest new particle $A_H$, the partner of the SM $U(1)_Y$ gauge boson,
becomes stable and weakly-interacting,
and thus a good dark matter candidate~\cite{LHT:DM}.
Many interesting phenomenological signatures at high energy colliders 
have been studied in the literature~\cite{Hubisz:pheno,Yuan:T:odd,WHWH:Cao,LHT:pheno},
including the implication for the LHC Higgs data~\cite{LHT:Higgs}.

The new heavy partner of the SM neutral $SU(2)$ gauge boson, $Z_H$,
is especially interesting since it decays almost exclusively into the SM Higgs boson
and the cold dark matter particle $A_H$.
By observing the Higgs boson with large $\met$,
we have an additional channel to probe the LHT model.
The $Z_H$ is mainly produced associated with
$W_H$~\cite{Yuan:T:odd}:
the $Z_H Z_H$ production cross section is about an order of magnitude smaller.
The $W_H$ boson decays into $W A_H$.
If $W$ decays leptonically, the final state of $ Z_H W_H$ production is 
$\ell\bb+\met$,
where two $A_H$'s and one neutrino carry $\met$.
We shall suggest optimal search strategies 
to enhance the signal significance at the LHC.

The rest of the paper is organized as follows.  
We begin our discussion with a brief review of the LHT model in Sec.~\ref{sec:review}.  
In Sec.~\ref{sec:production},
we specify the model parameters
for $Z_H W_H$ production,
and compute the total production cross sections
at the LHC.
We also suggest kinematic cuts to suppress the SM backgrounds,
and present the signal significance for the LHC energy of $\sqrt{s}=7,~8,~14\tev$.
We conclude in Sec.~\ref{sec:conclusions}. 

\section{Brief review of the LHT model}
\label{sec:review}
The LHT model is based on an $SU(5)/SO(5)$
non-linear $\sigma$-model with a gauge symmetry $[SU(2) \otimes U(1)]^2$.
The leading two-derivative term for the sigma field $\Sigma$ is
\beq
\label{eq:2dim}
\mathcal{L}  _{\Sigma} = \frac{1}{2} \frac{f^2}{4}
    {\rm Tr} | \mathcal{D}_{\mu} \Sigma |^2 ,
\eeq
where $f$ is the symmetry breaking scale of the order of 1 TeV.
The covariant derivative of
the sigma field is given by
\beq
\label{eq:covariantD}
\mathcal{D}_\mu \Sigma=  \partial_\mu\Sigma - i \sum_{j=1}^2\left\{
g_jW_j^a( Q_j^a \Sigma +  \Sigma Q_J^{aT})
+ g'_j B_j(Y_j\Sigma + \Sigma Y_j^T) \right\}.
\eeq
Here $B_j$ and $W^a_j$ are $U(1)_j$ and $SU(2)_j$ gauge fields,
respectively,
and their corresponding couplings are $g^\prime_j$ and $g_j$.
The generators are
$Q_1^a = \mathrm{diag}\left(
{\sigma^a}/{2} ,{\mathbf{0}}_{3\times 3}\right)$,
$Q_2^a = \mathrm{ diag}\left(
{\mathbf{0}}_{3\times 3} , -{\sigma^{a*}}/{2}
\right)$,
$Y_1 = \mathrm{diag}(-3,-3,2,2,2)/10$,
and $Y_2 = \mathrm{diag}(-2,-2,-2,3,3)/10$.

Symmetry breaking occurs in two stages:
(i)
the global $SU(5)$ symmetry as well as the $[SU(2) \otimes U(1)]^2$ gauge symmetry
are broken by non-zero vacuum expectation value (VEV) of an $SU(5)$ symmetric tensor $\Sigma_0$;
(ii) the electroweak symmetry is broken 
as the Higgs field develops non-zero VEV at loop levels.
The first stage symmetry breaking is from non-zero $\Sigma_0$ field given by
\begin{equation}
\label{Sigma0}
\Sigma_0 = \left( \begin{array}{ccc}
 & & {\mathbf{1}}_{2 \times 2} \\
 &1 & \\
{\mathbf{1}}_{2 \times 2} & &
\end{array}\right).
\end{equation}
This $\Sigma_0$ breaks the global $SU(5)$ symmetry 
into $SO(5)$ and the gauge symmetry
$[SU(2)\otimes U(1)]^2$ 
into its diagonal subgroup $SU(2)_L \otimes
U(1)_Y$.
Among 14 massless Goldstone
bosons (${\mathbf{1}}_0 \oplus {\mathbf{3}}_0 \oplus {\mathbf{2}}_{\pm \frac{1}{2}} 
\oplus {\mathbf{3}}_{\pm 1}$
representations of the $SU(2)_L$ gauge group)
from the global symmetry breaking,
${\mathbf{1}}_0$ and ${\mathbf{3}}_0$
are eaten by the broken gauge bosons $\vec{W}_H^{\mu }$ and $B_H^{\prime\mu }$.

The remained Goldstone degrees of freedom,
the $SU(2)_L$ doublet $h$ and triplet $\phi$, are parameterized by the pion matrix 
$\Pi$.
The low energy dynamics of the model is described by the expansion
of the sigma field as
\bea
\label{eq:expansion}
\Sigma = e^{2 i \Pi} \Sigma_0,
\eea
where the pion field is
\bea
\label{eq:Pi}
\Pi = \left( \begin{array}{ccccc}
\phi^{\dagger} & \frac{h^{\dagger}}{\sqrt{2}} & {\mathbf{0}}_{2\times
2} \\
\frac{h^{*}}{\sqrt{2}} & 0 & \frac{h}{\sqrt{2}} \\
{\mathbf{0}}_{2\times 2} & \frac{h^{T}}{\sqrt{2}} & \phi
\end{array} \right)
\,.
\eea

The second stage of symmetry breaking
occurs by the complex doublet
${\mathbf{2}}_{\pm \frac{1}{2}}$, which has proper quantum numbers for the SM Higgs boson.
Its interactions with gauge bosons and fermions
generate non-zero VEV as well as its mass at loop level.
The quadratic divergence in the radiative Higgs boson mass
is avoided by collective symmetry breaking.
If one set of $SU(2)\times U(1)$ gauge couplings vanishes,
the theory is invariant under a global $SU(3)$ symmetry.
The Higgs field remains as an exact Goldstone boson.
We need two symmetry breakings for the Higgs radiative mass,
which is log-divergent at one-loop level and quadratically divergent at two-loop level.
For $\Lm \sim \mathcal{O}(10)\tev$,
the Higgs boson mass is naturally 
of the order of 100 GeV. 

An inevitable consequence of non-zero VEV of the Higgs field is the tree level mixing between
the heavy new particles and the SM particles,
which is strongly constrained by the electroweak precision data~\cite{EWPD}.
Later $T$-parity was introduced in order to forbid the tree-level mixing,
under which the gauge bosons and pion field are transformed as
\bea
W_1^a \longleftrightarrow W_2^a,
\quad
B_1 \longleftrightarrow B_2,
\quad
\Pi \longleftrightarrow - \Omega \Pi \Omega,
\eea
where $\Omega = {\rm diag} (1,1,-1,1,1)$.
If we impose 
$g_1 = g_2 = \sqrt{2} g$ and
$g_1' = g_2' = \sqrt{2} g'$,
the Lagrangian in Eq.~(\ref{eq:covariantD}) is $T$-parity invariant.
The heavy gauge bosons are simply 
$W_H^\pm=  (W_1^\pm - W_2^\pm)/\sqrt{2}$, $Z_H = (W_1^3 - W_2^3)/\sqrt{2}$ and
$A_H = (B_1 - B_2)/\sqrt{2}$ with the masses of $M_{W_H} =M_{Z_H} = g f$ and 
$A_H = g'f/\sqrt{5}$ to leading order.
New gauge bosons of $W_H$, $Z_H$, and $A_H$
have
odd $T$-parity while all the SM fields have even $T$-parity.
There is no tree-level mixing between a SM particle and a new particle.

Another key ingredient in the LHT model is the presence of
$T$-odd fermions,
which is crucial for the invariance of $T$-parity in the model.
Their heavy masses are parameterized by Yukawa type coupling
$\kappa$ as $M_{f_-} = \sqrt{2} \kappa f$~\cite{Todd:fermion, Hubisz:pheno, Hubisz:EWPD}.
For simplicity, we assume that  $\kappa$ is flavor-diagonal and universal.
There are lower and upper bounds on $\kp$.
If $\kappa$ is too small,
new fermions become light and are to be copiously produced at the LHC.
The pair production of $T$-odd fermions leaves the final states of jets and missing transverse
energy, which is constrained by the search for squarks~\cite{LHC:squark}.
If $\kappa$ is too large, a naive expectation is that all heavy fermions
get decoupled from the theory.
However four fermion contact interactions have non-decoupling effects from
$\kappa$.
The $e^+ e^- \to q\bar{q}$ scattering constrains
$\kappa \lsim 3.4$~\cite{Hubisz:EWPD}.

Final comments are on the decay modes of the heavy gauge bosons of $Z_H$ and $W_H$.
Crucial are the two-body ferminonic decays of
$Z_H/W_H \to f_- f^{(\prime)}$.
For $\kp \geq 0.46$, however,
heavy masses of $f_-$'s close these decay modes kinematically.
The dominant decay mode of $Z_H$ is into $H A_H $.
The next dominant is into $H H A_H$ with the branching ratio of a few percent at most.
With $m_H=125\gev$, we have
$\br(Z_H \to A_H H) =99\%, ~97\%$
for $f =0.5, 1\tev$ respectively.  
The heavy $W_H$ decays almost exclusively into $W A_H$.

\section{$Z_H W_H$ signal at the LHC}
\label{sec:production}

\begin{figure}
\includegraphics[scale=1]{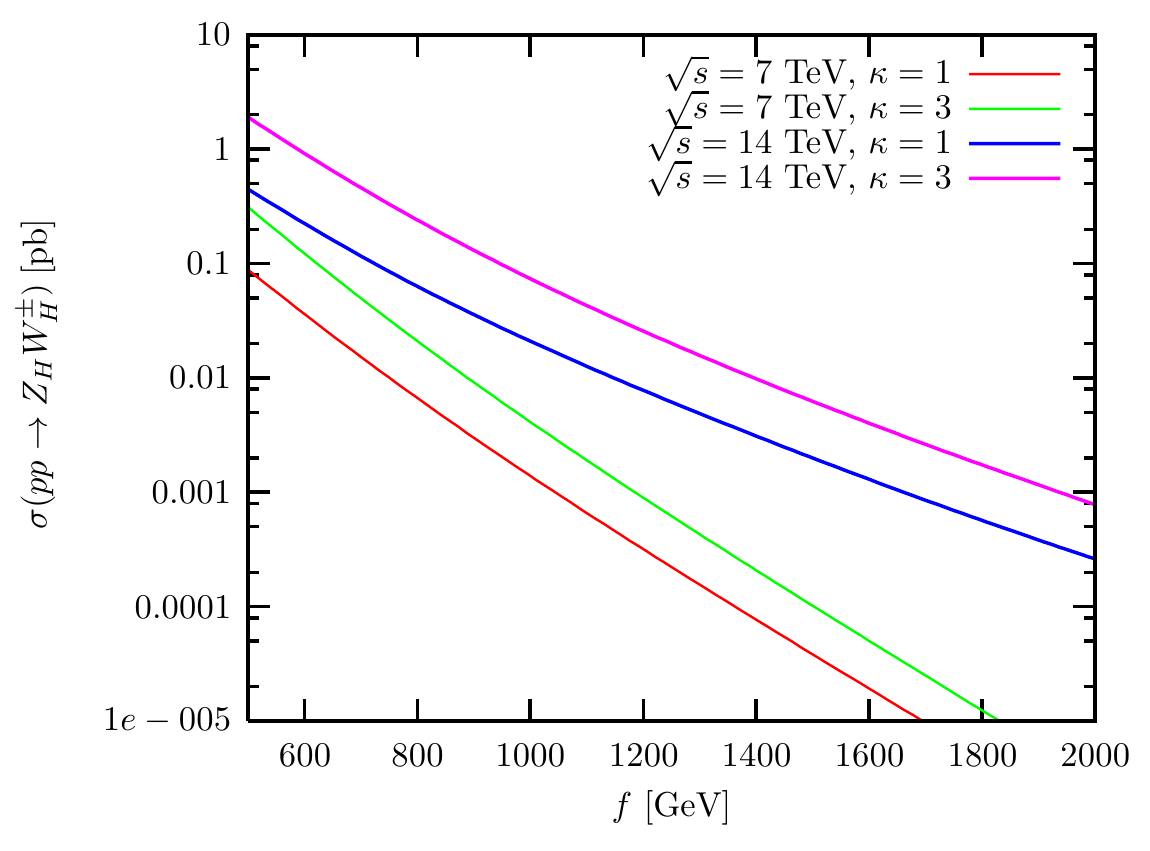}
\medskip
\caption{\small \label{fig:sigma}
The total cross section of $Z_H W_H$
at the LHC with $\sqrt{s}=7,14\tev$ for $\kappa=1,3$
as a function of $f$.
}
\end{figure}

The $Z_H W_H$ production 
at the LHC is from the $q \bar{q}'$ annihilation.
There are three Feynman diagrams: the 
$s$-channel one mediated by the SM $W$ boson,
the $t$-channel and $u$-channel ones by $T$-odd quarks.
An important observation is
that the $s$-channel contribution alone diverges
as energy goes to infinity~\cite{Yuan:T:odd}.
The $t$-channel and $u$-channel contributions interfere destructively
with the $s$-channel one, and 
cancel the diverging contributions.
The inclusion of $T$-odd heavy fermions 
is crucial in order to respect the partial-wave unitarity.

In Fig.~\ref{fig:sigma}, we show the total cross section of $Z_H W_H$
at the LHC with $\sqrt{s}=7,14\tev$ for $\kappa=1,3$
as a function of $f$.
The result for $\sqrt{s}=14\tev$ and $\kp=1$
is consistent with that in Ref.~\cite{Yuan:T:odd}.
If $\kappa=1$ and $f=500\gev$,
the total cross section is about 0.45 pb (0.09 pb)
for $\sqrt{s}=14 \;(7)\tev$.
If $\kappa$ increases into 3,
the cross section also increases
by a factor of $3 \sim 4$.
This is attributed to the weakened destructive interference effects.
At the LHC with $\sqrt{s}=8\tev$,
the total production cross section,
compared to $\sqrt{s}= 7 \tev$, increases by $6-7\%$.
Recently the QCD $K$-factor at next-to leading order
has been calculated to be roughly $\sim 1.1$~\cite{K-factor}.

\begin{figure}[t!]
\includegraphics[scale=1]{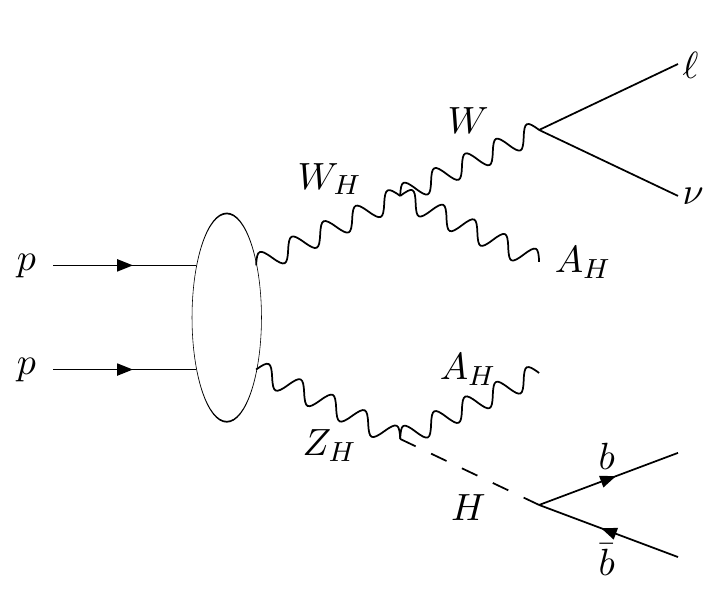}
\medskip
\caption{\small \label{fig:lbb}
$\ell b \bar{b} +\met$ channel from the $Z_H W_H$ decay
}
\end{figure}

We consider the dominant subsequent decays 
of $Z_H W_H$ as in Fig.~\ref{fig:lbb}:
\bea
\label{eq:process}
p p \to Z_H + W_H \to H A_H  + W A_H \to b\bar{b} A_H + \ell\nu A_H.
\eea
Note that both the branching ratios of $\br(Z_H \to A_H H)$ and $\br(W_H \to W A_H)$
are almost 100\%
for $\kp=1,3$.
We use MADGRAPH~\cite{MADGRAPH} for the signal event generator.
The basic cuts on the transverse momentum 
$p_T$ and the pseudo-rapidity $\eta$ of jets and charged leptons
are
\bea
\label{eq:basic:cut}
%\hbox{\sffamily Basic Cuts: }&&
p_{Tj} > 50\gev,
\quad  p_{T \ell} > 25\gev,
 \quad
|\eta_{j,\ell}| < 2.5.
\eea

The irreducible SM backgrounds are 
\bea
\label{eq:bbl:Wbb}
pp &&\to W + b\bar{b}\to \ell \nu + b\bar{b}, ~\hbox{(QCD)}\\
\label{eq:bbl:WZ}
pp   && \to W + Z \to \ell \nu + b\bar{b}, \\
\label{eq:bbl:WH}
pp   && \to W + H \to \ell \nu + b\bar{b}.
\eea
Dominant is the first QCD process in Eq.~(\ref{eq:bbl:Wbb}), 
mediated by the gluon.
The cross section of this QCD process with the basic cuts in Eq.~(\ref{eq:basic:cut})
is $\sim \mathcal{O}(20)\pb$.
Our signal with the basic cuts is $\sim \mathcal{O}(1)\fb$.
Other backgrounds are $pp \to WZ$ and $pp \to WH$.
A reducible background comes from $\ttb$ production followed by $t\bar t$
decaying into
$b\bar{b}\ell\ell\nu\nu$ with one of the charged leptons missed in the 
detector.

Now we suggest suitable kinematic cuts
to suppress the SM backgrounds.
First we note that the $\bb$ in our signal is from the decay of the Higgs boson,
while the $\bb$ in the QCD background is mediated by a gluon.
We apply the following cut on the invariant mass of $\bb$
to suppress the SM backgrounds, especially the QCD one:
\bea
\label{eq:Mbb:cut}
%\hbox{\sffamily $M_{\bb}$ Cut: }
115\gev < M_\bb < 135 \gev.
\eea

Another sensitive probe is the missing transverse energy.
The SM background has $\met$ from the neutrino in the $W$ decay,
while the $\met$ in our signal comes from two heavy $A_H$'s and  
one neutrino.
Naturally
our signal has a harder $\met$ distribution.
We apply the following  strong $\met$ cut:
\bea
\label{eq:MET:cut}
%\hbox{\sffamily $\met$ Cut}: 
\met > 100 \gev.
\eea
Finally we apply $b$-tagging for both $b$ quarks with the following
efficiency~\cite{b-tag}:
\bea
\epsilon_b = 67\%,
\quad 
\epsilon_{\rm mistag} =1.5\%.
\eea

%Finally the reducible $\tt$ backgrounds with one lepton missing are to be suppressed by imposing
%\bea
%\label{eq:tt:cut}
%\hbox{\sffamily $\tt$ Cuts: } &&
%\Delta R_{bb} < XXX , \quad \Delta R_{\ell b} > XXX,
%\eea
%where $\Delta R_{ij} =\sqrt{ (\Delta \eta_{ij})^2+ (\Delta \phi_{ij})^2}$.
%This cut is to 
%The $b$ quark pair decay from the Higgs boson tend to 

\begin{table}[tb!]
\centering
\caption{\small \label{table:bbl}
Signal and background cross sections as well as the signal significance
for $pp \to \ell\bb+\met$
at the LHC  with the total luminosity ${\cal L}=11.2\ifb$ at $\sqrt{s}=7\tev$,
${\cal L}=40\ifb$ at $\sqrt{s}=8\tev$,
and ${\cal L}=100\ifb$ at $\sqrt{s}=14\tev$.
Additional kinematics cuts are 
$115< M_{\bb}<135\gev$ and $\met>100\gev$.
We have applied 10\% systematic uncertainty,
the $b$-tagging efficiency of $\epsilon_b =67\%$,
and the mistagging efficiency of $\epsilon_{\rm mistag}=1.5\%$.
}
\begin{ruledtabular}
\begin{tabular}{c|cc|cc|cc}
& \multicolumn{6}{c}{$\ell\bb+\met$ ($115< M_{\bb}<135\gev$ and $\met>100\gev$)} \\ \hline
$\sqrt{s}$ at the LHC  & \multicolumn{2}{c|}{ $7\tev$} 
	& \multicolumn{2}{c|}{$8\tev$} 	& \multicolumn{2}{c}{$14\tev$} \\ \hline
$\sg_{\rm SM}$ & \multicolumn{2}{c|}{1.8 fb}  & \multicolumn{2}{c|}{2.2 fb}
 & \multicolumn{2}{c}{5.0 fb} \\ \hline
	& $\sg_{\rm signal}$ & $\frac{S}{\sqrt{B}\oplus 0.1 B}$
	& $\sg_{\rm signal}$ & $\frac{S}{\sqrt{B}\oplus 0.1 B}$
	& $\sg_{\rm signal}$ & $\frac{S}{\sqrt{B}\oplus 0.1 B}$ \\[0.5ex] \hline
	$\kp=1$  & \multirow{2}{*}{ 2.0 fb }  & \multirow{2}{*}{ 3.5 } 
& \multirow{2}{*}{ 2.9 fb }  & \multirow{2}{*}{ 6.3 } 
& \multirow{2}{*}{ 9.9 fb }  &  \multirow{2}{*}{ 13.7 }  
 \\
$f=500\gev$ & & & & & & \\ \hline
%---------
$\kp=3$ & \multirow{2}{*}{ 7.4 fb } & \multirow{2}{*}{ 12.9 } 
& \multirow{2}{*}{ 11.1 fb }  & \multirow{2}{*}{ 23.4 } 
& \multirow{2}{*}{ 45.0 fb }  &  \multirow{2}{*}{ 62.0 }  \\
$f=500\gev$ & & & & & & \\ \hline
%---------
$\kp=3$ & \multirow{2}{*}{ 1.1 fb } & \multirow{2}{*}{ 1.9 } 
& \multirow{2}{*}{ 1.8 fb }  & \multirow{2}{*}{ 4.0 } 
& \multirow{2}{*}{ 10.0 fb }  &  \multirow{2}{*}{ 13.7 }  \\
$f=700\gev$ & & & & & & \\ \hline
%---------
$\kp=3$ & \multirow{2}{*}{ 0.1 fb } & \multirow{2}{*}{ 0.2 } 
& \multirow{2}{*}{ 0.2 fb }  & \multirow{2}{*}{ 0.5 } 
& \multirow{2}{*}{ 1.8 fb }  &  \multirow{2}{*}{ 2.5 }  \\
$f=1\tev$  & & & & & & \\ 
\end{tabular}
\end{ruledtabular}
\end{table}

In Table \ref{table:bbl}, we summarize
the signal and SM background cross sections,
and the significance 
for $pp \to \ell\bb+\met$
at the LHC  with the total luminosity ${\cal L}=11.2\ifb$ at $\sqrt{s}=7\tev$,
${\cal L}=40\ifb$ at $\sqrt{s}=8\tev$,
and ${\cal L}=100\ifb$ at $\sqrt{s}=14\tev$.
For the model parameters, we consider four
cases of $(f/\tev,\kp)=(0.5,1),~(0.5,3),~(0.7,3),~(1,3)$.
We have applied all of the above cuts.
We also include 10\% systematic uncertainty when computing the significance,
which is inevitable with $b$ jets.

First we compare two cases of $\kp=1$ and $\kp=3$ with the given $f=500\gev$.
A larger $\kp$ value, which implies heavier $T$-odd quarks and smaller destructive interference
between $s$-channel and $t(u)$-channel diagrams,
leads to more signal events:
the cross section of the $\kp=3$ case is bigger than that of the $\kp=1$ case
by a factor of $3.7 \sim 4.5$.
For $f=500\gev$ even with small $\kp$,
the LHC has great potential of discovering
$Z_H W_H$ production through the final state of $\ell\bb+\met$, 
especially at $\sqrt{s}=8,14\tev$.
Increasing the value of $f$ will lead to smaller cross sections.
If $f=700\gev$ and $\kp=3$,
the discovery significance is high enough about 4 at $\sqrt{s}=8 \tev$
and above 13 at $\sqrt{s}=14\tev$.
Although
it is reduced considerably compared to 
the $f=500\gev$ case, this case can be probed at the LHC.
If $f=1\tev$,
the decrease of the cross section is too severe.
Even with a high luminosity $100\ifb$ at $\sqrt{s}=14\tev$ and 
even in the large $\kp$ case, the result
does not yield high enough significance.
The discovery significance of 5 at $\sqrt{s}=14\tev$ is obtained if
$f\lsim 850 \gev$. 

\section{Conclusions}
\label{sec:conclusions}
Upon the discovery of a new boson with mass around 125 GeV, 
which shows incredible resemblance with the SM Higgs boson,
we have studied 
the potential of the Higgs boson as a tagging particle in the LHT model.
The $ Z_H W_H$ production channel at the LHC
has several merits for this purpose:
(i) it has the largest cross section among heavy gauge boson production;
(ii) both $W_H$ and $Z_H$ have practically one decay mode
of $W_H \to W A_H$ and $Z_H \to H A_H$; 
(iii) the subsequent decays lead to a 
relatively clean final state of $\ell\bb+\met$.

We have shown that the $\ell\bb+\met$ final state from
the $ Z_H W_H$ production channel can yield high enough
discovery significance for testing the model.
With suggested kinematic cuts we are able to reduce the SM backgrounds
to a manageable level, including the large QCD $Wb\bar b$,
$WZ$, and $WH$ production, as well as the reducible background of
$t\bar t$ production.
The signal significance at the LHC with $\sqrt{s}=14\tev$
can be risen above 5 if $f \lsim 850\gev$ and $\kp=3$.

\acknowledgments
This work is supported by WCU program through the KOSEF funded
by the MEST (R31-2008-000-10057-0) and by the 
the National Science Council of Taiwan under Grants No.~99-2112-M-007-005-MY3.
KYL is supported by the Basic Science Research Program
through the NRF funded by the MEST (2010-0010916).

%%%%%%%%%%%%%%%%%% References
%%%%%%%%%%%%%%%%%%%%%%%%%%%%%%%%%%%%%%%%%%%%%%%%%%%%%%%
\def\PRD #1 #2 #3 {Phys. Rev. D {\bf#1},\ #2 (#3)}
\def\PRL #1 #2 #3 {Phys. Rev. Lett. {\bf#1},\ #2 (#3)}
\def\PLB #1 #2 #3 {Phys. Lett. B {\bf#1},\ #2 (#3)}
\def\NPB #1 #2 #3 {Nucl. Phys. B {\bf #1},\ #2 (#3)}
\def\ZPC #1 #2 #3 {Z. Phys. C {\bf#1},\ #2 (#3)}
\def\EPJ #1 #2 #3 {Euro. Phys. J. C {\bf#1},\ #2 (#3)}
\def\JHEP #1 #2 #3 {J. High Energy Energy Phys. #1 (#2) #3}
\def\IJMP #1 #2 #3 {Int. J. Mod. Phys. A {\bf#1},\ #2 (#3)}
\def\MPL #1 #2 #3 {Mod. Phys. Lett. A {\bf#1},\ #2 (#3)}
\def\PTP #1 #2 #3 {Prog. Theor. Phys. {\bf#1},\ #2 (#3)}
\def\PR #1 #2 #3 {Phys. Rep. {\bf#1},\ #2 (#3)}
\def\RMP #1 #2 #3 {Rev. Mod. Phys. {\bf#1},\ #2 (#3)}
\def\PRold #1 #2 #3 {Phys. Rev. {\bf#1},\ #2 (#3)}
\def\IBID #1 #2 #3 {{\it ibid.} {\bf#1},\ #2 (#3)}

%%%%%%%%%%%%%%%%%%%%%%%%%%%%%%%%%%%%%%%%%%%%%%%%%%%%%%%%%%

\end{document}